\documentclass[runningheads]{llncs}
\usepackage{graphicx}
\usepackage{hyperref}
\usepackage{latexsym}
\usepackage{booktabs}
\usepackage{todonotes}
\usepackage[hang,flushmargin]{footmisc}
\usepackage{colortbl}
\usepackage{amsfonts}
\usepackage{enumitem}
\usepackage[normalem]{ulem}
\usepackage{setspace}

\begin{document}
\title{Ranking Significant Discrepancies \\ in Clinical Reports}
\titlerunning{Ranking Significant Discrepancies in Clinical Reports}

\author{
Sean MacAvaney\inst{1} \and
Arman Cohan\inst{2} \and
Nazli Goharian\inst{1} \and
Ross Filice\inst{3}}

\institute{
{IR Lab, Georgetown University, Washington DC, USA \email{\{firstname\}@ir.cs.georgetown.edu}} \and
{Allen Institute for Artificial Intelligence, Seattle, WA, USA \\ \email{armanc@allenai.org}} \and
{Dept. of Radiology, MedStar Georgetown Univ. Hospital, Washington DC, USA \\ \email{Ross.W.Filice@medstar.net}}
}
\authorrunning{S. MacAvaney et al.}

\maketitle              % typeset the header of the contribution
\begin{abstract}
Medical errors are a major public health concern and a leading cause of death worldwide. 
% It is both challenging to uncover consistent causes of errors and even when identified, to provide a solution that minimizes the chances of recurrent errors. 
Many healthcare centers and hospitals use reporting systems where medical practitioners write a preliminary medical report and the report is later reviewed, revised, and finalized by a more experienced physician. The revisions range from stylistic to corrections of critical errors or misinterpretations of the case. Due to the large quantity of reports written daily, it is often difficult to manually and thoroughly review all the finalized reports to find such errors and learn from them. To address this challenge, we propose a novel ranking approach, consisting of textual and ontological overlaps between the preliminary and final versions of reports. The approach learns to rank the reports based on the degree of discrepancy between the versions. This allows medical practitioners to easily identify and learn from the reports in which their interpretation most substantially differed from that of the attending physician (who finalized the report). This is a crucial step towards uncovering potential errors and helping medical practitioners to learn from such errors, thus improving patient-care in the long run.
We evaluate our model on a dataset of radiology reports and show that our approach outperforms both previously-proposed approaches and more recent language models by 4.5\% to 15.4\%.

\end{abstract}

\section{Introduction}\label{sec:intro}

Medical errors are a pervasive problem in healthcare that can result in serious patient harm~\cite{makary2016medical}. To identify and reduce the occurrence of preventable errors, many medical centers use reporting systems to document cases. Initial reports are often reviewed and revised by more experienced physicians. The revisions could be due to stylistic reasons or (more importantly) misinterpretations/errors in the initial report. In such cases, to prevent recurrence of the errors, is crucial to identify reports with substantive differences between the original and final report and discuss them with the clinician who wrote the initial report. 
It is often challenging to manually identify such cases among the large number of daily written reports in a timely manner. In this work, we propose an approach for ranking revisions of medical reports by the degree of discrepancy between the different versions of the report. This allows medical practitioners to easily find the reports in which they made an error, which helps them learn from their mistakes and prevent future similar errors.

This is a challenging task to automate because the edits that an attending physician makes to a report can range from stylistic differences to significant discrepancies that may have a major effect on the patient (e.g., an unobserved mass). See Figure~\ref{fig:exampledis} for an example of significant and non-significant discrepancies from radiology reports. As we can see, differences between the significant and non-significant discrepancies are often not trivial to identify and requires more than just comparing surface word changes in the versions of the reports. Furthermore, significant discrepancies can occur relatively frequently in practice; in our dataset collected from a large urban hospital, around 7\% of reports contained significant errors. With hundreds of reports generated a week at some hospitals, this can amount to a considerable number of errors. 
We address this problem by proposing a supervised ranking approach for clinician's revised reports by the degree that there are significant discrepancies between their preliminary report and the final corrected report. I.e., our goal is to rank revisions that are more likely due to errors higher than revisions that are merely due to stylistic changes.

\begin{figure}[t]
\centering
\fbox{\parbox{2.2in}{\scriptsize
...with imaging features strongly suggestive of hepatocellular carcinoma (LI-RADS 4) \sout{not well discernible} \underline{probably present but not conspicuous} on prior examination...
}}
\fbox{\parbox{2.2in}{\scriptsize
... 3. Left renal artery: Single with \sout{a} \underline{slightly early branching first} branch point \sout{averaging} \underline{averages} 1.9 cm from the left lateral margin of the aorta. Left renal vein: Single without late confluence...
}}
\parbox{2.2in}{\centering\scriptsize\bf Significant discrepancy}
\parbox{2.2in}{\centering\scriptsize\bf Non-Significant discrepancy}
\vspace{-0.5em}
\caption{Example radiology impression revisions (strikeout removed, underline added). We aim to rank report revisions by the significance of the discrepancy.}\vspace{-1.2em}
\label{fig:exampledis}
\end{figure}

Prior works have investigated significant discrepancies in medical reports through comparison of surface textual features \cite{walls2009depictors,ruutiainen2011identifying}, semantic similarity features \cite{cohan2016identifying}, and word frequencies \cite{kalaria2015comparison}. These works often treat the problem as classification and the most successful ones leverage a variety of textual similarity measures. Viewing this problem as ranking is a more suitable and practical form of evaluation; given a doctor's limited time, it is important for them to be presented with the reports that have the most significant discrepancies. 

Document ranking in the broad medical domain have received extensive interest of researchers \cite{Yates2014RelevanceRankedDS,roberts2017overview,Soldaini2017DenoisingCN,sankhavara2018biomedical,Koopman2017GeneratingCQ,Saleh2019TermSF}.
However, these efforts focus on conventional query-document retrieval. Our goal is to rank significant discrepancies by measuring the semantic overlap between the initial and final report. 
There have also been efforts to identify semantic similarity between two texts, e.g., for paraphrase identification~\cite{Gan2016LearningGS,Tien2018SentenceMV,Liu2019IncorporatingCA,Reimers2019SentenceBERTSE}, but these approaches operate on the sentence-level, making them unsuitable for documents (e.g., radiology reports).

To summarize, our contributions are: (i) we propose an end-to-end supervised ranking model for identifying significant discrepancies in medical reports.
(ii) We demonstrate that our approach outperforms both previously proposed approaches and more recent language model approaches in a variety of metrics. 
(iii) We provide an analyses of the importance of different model components.

\section{Model}\label{sec:methods}

We propose a supervised model that measures the overlap between the preliminary and final report for the purpose of ranking pairs of preliminary and final reports based on their significance over a given period of time.
We observe that the central challenge of this task is being permissive of surface-level changes (which may be considerable), while emphasizing changes of substance, which may be subtle (see Figure~\ref{fig:exampledis} for examples of such changes). To address this, we incorporate \textit{importance} and \textit{similarity} scores. The \textit{importance score} weights each term/phrase and is learned during training. This score allows for terms that are not important to have less of an impact on the ranking score of the report (e.g., words like \textit{well} and \textit{but}). 
Note that this is a special application in which some function words that are often ignored actually have a big impact on the meaning of a report (e.g., \textit{not} is often considered a stop word and removed).
We let the model learn which terms are important during training. The \textit{matching score} allows for the model to account for the replacement of similar terms using the cosine distance of word vectors (e.g., \textit{averaging} and \textit{averages} are similar) and synonym information from a domain-specific ontology (\textit{chauffeur fracture} and \textit{Hutchinson fracture} are synonymous). This allows the replacement of semantically-similar terms to have little impact on the ranking score. We calculate three \textit{similarity scores} (addition, deletion, and overlap) using the importance and matching scores, and linearly combine them as a \textit{ranking score}.

\textbf{Notation and task definition.} Let $R$ be a set of clinical reports. Each report $r\in R$ consists of a preliminary and final version of the report ($p$ and $f$, respectively), and a label $l\in\{0,1,...,L\}$ indicating the degree of discrepancy between $p$ and $f$. Each version of the report consists of a sequence of tokens, denoted by $p_i$ and $f_i$. The significant discrepancy ranking task produces a ranking score $s\in \mathbb{R}$ for each report $r\in R$ such that the reports with higher degrees of discrepancy are assigned a higher ranking score. % (i.e., appear higher in the ranked list).

\textbf{Similarity scores.} Our approach combines several similarity scores to produce a ranking score. Specifically, we measure the weighted soft additions, deletions, and overlap of unigrams, n-grams, and ontological entities. The addition score ($S_a$, Eq.~\ref{eq:add}) defines weighted soft similarity as the ratio between the similarity score (weighted by a learned importance score) and the total importance of all terms in the final report. Thus, terms from the final report that do not appear in the preliminary report (i.e., additions) yield a higher score.
The deletion score ($S_d$, Eq.~\ref{eq:del}) is defined similarly, but in terms of the preliminary report; terms from the preliminary report that do not appear in the final report (deletions) yield a higher score. The overlap score ($S_o$, Eq.~\ref{eq:ovelrap}) combines the addition and deletion scores into one succinct measure. We use all three scores to measure term unigram, n-gram, and ontological differences (defined below). We define the similarity functions (where $M_X(y)\in[0,1]$ is a matching score of term $y$ in $X$, and $I(y)\in[0,1]$ is the importance score of term $y$) as:
\vspace{-2em}
\begin{multicols}{2}
\begin{equation}\scriptsize\label{eq:add}
S_{a}(p,f)=-\frac{\sum_{f_i\in f} M_p(f_i)I(f_i)}{\sum_{f_i\in f} I(f_i)}
\end{equation}\break
\begin{equation}\scriptsize\label{eq:del}
S_{d}(p,f)=-\frac{\sum_{p_i\in p} M_f(p_i)I(p_i)}{\sum_{p_i\in p} I(p_i)}
\end{equation}
\end{multicols}
\begin{equation}\scriptsize\label{eq:ovelrap}
S_{o}(p,f)=-\frac{\sum_{p_i\in p} M_f(p_i)I(p_i) + \sum_{f_i\in f} M_p(f_i)I(f_i)}{\sum_{p_i\in p} I(p_i) + \sum_{f_i\in f} I(f_i)}
\end{equation}

\textbf{Unigram and n-gram matching.} Unigram matching can provide valuable signals for significance in radiology reports. For instance, the addition \textit{no} (e.g., \textit{fracture} vs. \textit{no fracture}) could change the meaning of the report considerably. We define the matching function for unigrams as the maximum cosine similarity between the word embeddings ($emb(\cdot)$) of the term and any term in the other report, and a unigram importance function using a simple feed-forward layer with sigmoid activation ($W_{imp}$ and $b_{imp}$ as model parameters):
\vspace{-2em}
\begin{multicols}{2}
\begin{equation}\small
M_X(y)=\max_{x\in X}(\cos(emb(x), emb(y)))
\end{equation}\break
\begin{equation}\small
I(y)=\sigma(emb(y)W_{imp}+b_{imp})
\end{equation}
\end{multicols}

N-gram matching provides another important view of similarity, since there are many multi-word noun phrases in radiological notes. For instance, \textit{right arm} and \textit{left arm} represent completely different parts of the body, and should be treated differently. We handle n-grams by first taking the average of the embeddings over sliding windows. This is a simple and effective way to combine the representations. We use bi-grams and tri-grams in our experiments.

\textbf{Ontological matching.} Since medical knowledge is broad and extensive, the model may never encounter certain medical entities during training. This knowledge may also not be captured effectively by embeddings. Thus, it is valuable to explicitly encode domain information into the model using an ontology. We use a mapping function that matches any exact ontological name to the corresponding concept, a constant similarity for exact entity matches, and constant weight for all ontology concepts.
We use RadLex (v4.0, \url{http://radlex.org/}), an ontology of radiology concepts (e.g., procedures, diagnoses, etc.).

\section{Experiments}\label{sec:exp}

\textbf{Dataset.} We train and evaluate using a dataset of 3,368 radiology reports from a large urban hospital. Each sample consists of a preliminary report written by a resident, and a final report revised by the attending radiologist, who labeled the edit by the degree of discrepancy between the two reports. The labels are \texttt{0} (attending doctor fully agrees with assessment of the resident, 81\% of reports), \texttt{1} (errors exist, but they are insignificant to the overall impression, 12\%), \texttt{2} (subtle, yet important, error exists, 6\%), \texttt{3} (an obvious error exists, 1\%). We split the dataset into 122 sets based on the combination of resident and week (\textit{ranking sets}, average 27.6 reports per ranking set, min 5, max 148). Since residents often work weekly shifts, this is a valuable setting because it allows residents to review report discrepancies from the past week. We randomly split the ranking sets into 60-20-20 train-dev-test set splits. Each ranking set consists of at least 5 reports, each with at least one report discrepancy. Radiology reports contain several sections; we primarily concern ourselves with the summary section of the reports (called the \textit{impression}) because it contains the main findings.

\textbf{Baselines.} To evaluate the effectiveness of the model, we compare with the variety of methods in the state-of-the-art, including ranking models, domain specific models and textual similarity models, briefly described below:

\begin{itemize}[noitemsep,topsep=0pt,leftmargin=6pt]

\item[-] \textbf{Vector space model (VSM)}. We use the traditional TFIDF-weighted vector space similarity score between the preliminary and final report (from lucene).

\item[-] \textbf{BiPACRR}. We test the PACRR~\cite{Hui2017PACRRAP} neural IR model because it learns to identify n-gram similarity between two texts. We modify the architecture to learn two scores (one with the preliminary report as the query and the other with the final report as the query), and linearly combine them to produce a final ranking score. We call this variant BiPACRR. We also experimented with other neural rankers (e.g., KNRM~\cite{Xiong2017EndtoEndNA}), but BiPACRR was the most effective.

\item[-] \textbf{Textual similarity regression (SimReg)}~\cite{cohan2016identifying}. This approach uses logistic regression to combine several hand-crafted features (mostly consisting of textual similarity measures and lexical features) to identify significant discrepancies in radiology reports. Since this approach performs classification, we use the label score as the ranking score. Our experiments used the authors' implementation.

\item[-] \textbf{(Sci)BERT classification}. We use the standard fine-tuned BERT textual similarity method on both the pretrained BERT~\cite{Devlin2018BERTPO} (\texttt{base-uncased}) and SciBERT~\cite{Beltagy2019SciBERTPC} (\texttt{scivocab-uncased}) models. Based on preliminary parameter tuning, we use a learning rate of $10^{-5}$ for fine-tuning these models.

\end{itemize}

\textbf{Evaluation metrics.} Given the time constraints of doctors, we choose evaluation metrics that emphasize placing reports with higher discrepancies at the top. We evaluate using nDCG@1, nDCG@5, nDCG (without cutoff), P@1, P@5, and R-Prec (binary labels test any degree of discrepancy higher than label \texttt{0}).

\textbf{Parameters and training.}
We train the neural models using pairwise cross-entropy loss~\cite{Dehghani2017NeuralRM}. Hyper-parameters are tuned using nCDG@5 on the dev set. We use SciBERT term embeddings~\cite{Beltagy2019SciBERTPC} in our model and BiPACRR and tune for the optimal layer's embeddings akin to~\cite{macavaney:sigir2019-contextuallms}. SciBERT is an adaptation of BERT to the biomedical and scientific domains, making it suitable for radiology notes.

\textbf{Results.} Test set performance of our best model configuration are shown in Table~\ref{tab:main_results}. Our optimal model consists of unigram, bi-gram, tri-gram, and RadLex scores. When compared to the best prior work (SimReg~\cite{cohan2016identifying}), our model typically yields a considerable improvement in ranking performance. Our method improves R-Prec by 7.4\%, nDCG@1 and nDCG@5 performance by 4.5\%, and P@5 performance by 4.7\%. In 54\% of the test cases, our approach improves the nDCG@5 score over SimReg (decreases performance in only 27\% of cases). Our model also outperforms leading language model classification approaches (BERT and SciBERT) and a leading neural ranking approach tuned for this task in most metrics (BiPACRR) by up to 15.4\% in nDCG@1.
We attribute this improved effectiveness of our approach to the explicit modeling of term importance and overlap, which are critical for the task.

\begin{table}[t]
\caption{Ranking performance of our method and baselines.}
\vspace{-0.7em}
\label{tab:main_results}
\centering\small
\scalebox{0.8}{
\setlength{\tabcolsep}{7pt}
\begin{tabular}{@{}lrrrrrr@{}}\toprule
Model & nDCG@1 & nDCG@5 & nDCG & P@1 & P@5 & R-Prec \\
\midrule
VSM & 48.1 & 54.0 & 70.9 & 65.4 & 42.3 & 49.4 \\
BERT & 59.0 & 69.8 & 78.7 & 69.2 & 53.8 & 53.9 \\
SciBERT & 62.2 & 68.2 & 79.3 & 76.9 & 51.5 & 58.3 \\
BiPACRR & 64.1 & 68.6 & 77.5 & 69.2 & \bf56.2 & 55.3 \\
SimReg & 69.9 & 70.7 & 81.1 & \bf80.8 & 51.5 & 51.8 \\
Our Method & \bf74.4 & \bf75.2 & \bf83.7 &\bf80.8 &\bf56.2 & \bf59.2 \\
\bottomrule
\end{tabular}
}

\end{table}

\textbf{Ablations.} Table~\ref{tab:ablation} shows the ablation study examining the importance of different components in our system. We observe that both contextualization and domain-specificity of the word embeddings improve the performance of our approach. The term importance mechanism improves nDCG@5 by 6.9\% and the ontology similarity improves performance by 9.8\%. All three similarity measures appear to be important, however the overlap score alone can account for most of the performance (last row in table). This may be because it succinctly accounts for both additions and deletions.

\begin{table}[t]
\caption{Ablation study of our method.}
\vspace{-0.7em}
\label{tab:ablation}
\centering\small
\scalebox{0.8}{
\begin{tabular}{@{}lr@{}}\toprule
Model & nDCG@5 \\
\midrule

Full Model & \bf75.2 \\
 - replace SciBERT with BERT & 64.7 \\
 - replace SciBERT with BioNLP (\texttt{pubmed-pmc}, \url{bio.nlplab.org}) & 62.2 \\
 - replace SciBERT with FastText (\texttt{wiki-news-300d-1M}, \url{fasttext.cc}) & 59.1 \\
 - without term importance & 68.3 \\
 - without ontology similarity & 65.4 \\
 - only overlap score ($S_o$) & 70.8 \\
 - only addition/deletion scores ($S_a$ and $S_d$) & 61.5 \\

\bottomrule
\end{tabular}
}
\vspace{-6pt}
\end{table}

\textbf{Term importance.} To better understand the term importance mechanism of our approach, we present an example report in Figure~\ref{fig:importance} (slightly altered for privacy). This report contains highly significant discrepancies and was ranked at position 3 by our approach and position 9 by SimReg (below several non-significant discrepancies). We observe that our model considers many radiological conditions as important, both when unmodified between the reports and when added/deleted (e.g., \textit{fracture}, \textit{dislocation}, \textit{bankart}). Judging by the low textual similarity in this example, we conclude that the SimReg model may be relying too heavily on lexical features. We check the terms that are assigned high importance scores across all reports and find the most common are \textit{no} (12\% of reports), \textit{cardiopulmonary} (3\%), \textit{process} (3\%), and \textit{abnormality} (3\%).

\vspace{-1.5em}
\begin{figure}\small\centering
\scalebox{0.8}{
\fbox{\parbox{4.7in}{
\begingroup
\setlength{\fboxsep}{0pt}\small

\colorbox{blue!0}{\sout{anter}}\colorbox{blue!1}{\sout{oin}}\colorbox{blue!1}{\sout{fer}}\colorbox{blue!2}{\sout{ior}} \colorbox{blue!0}{\sout{dislocation}} \colorbox{blue!3}{\sout{of}} \colorbox{blue!4}{\sout{the}} \colorbox{blue!1}{\sout{left}} \colorbox{blue!2}{\sout{shoulder}}\colorbox{blue!5}{\sout{.}} \colorbox{blue!2}{\sout{mild}} \colorbox{blue!18}{\sout{hill}}\colorbox{blue!42}{\sout{-}} \colorbox{blue!10}{\sout{sac}}\colorbox{blue!8}{\sout{hs}} \colorbox{blue!2}{\sout{deformity}} \colorbox{blue!2}{\sout{without}} \colorbox{blue!3}{\sout{associated}} \colorbox{blue!8}{\sout{bank}}\colorbox{blue!9}{\sout{art}} \colorbox{blue!6}{\sout{lesion}}\colorbox{blue!2}{\sout{.}}
\colorbox{blue!0}{no} \colorbox{blue!0}{\underline{evidence}} \colorbox{blue!4}{\underline{of}} \colorbox{blue!0}{acute} \colorbox{blue!11}{fracture} \colorbox{blue!1}{\underline{or}} \colorbox{blue!16}{\underline{dislocation}} \colorbox{blue!1}{\sout{of}} \colorbox{blue!1}{\sout{the}} \colorbox{blue!0}{\sout{hum}}\colorbox{blue!0}{\sout{er}}\colorbox{blue!0}{\sout{us}}\colorbox{blue!1}{.}

\endgroup
}}
}
\vspace{-0.5em}
\caption{Example unigram importance scores (mean of preliminary and final report). Darker colors indicate higher scores. Underlines: additions. Strikeouts: deletions.}
\label{fig:importance}
\vspace{-0.8em}
\end{figure}

\vspace{-1em}
\noindent\textbf{Conclusions.}\label{sec:concl}
We presented a supervised ranking model based on lexical and ontological overlaps to rank medical reports by their discrepancy significance.
On a real-world dataset of medical reports, we demonstrated that our approach outperforms existing approaches by large margins. 
This direction is a critical step towards addressing the problem of medical errors. By allowing medical practitioners to more easily find and learn from their previous errors, the chance of recurrent errors will be reduced, improving the well-being of patients.

\vspace{0.5em}
\noindent\textbf{Acknowledgements.} This work was supported in part by ARCS Foundation.

\bibliographystyle{splncs04}
\bibliography{biblio}

\end{document}